# On the Possibility of an 'Astronomical Perspective' in the Study of Human Evolution[1]

**Elio Antonello**
INAF – Osservatorio Astronomico di Brera
Via E. Bianchi 46, 23807 Merate, Italy
elio.antonello@brera.inaf.it

**Abstract**
The 'Sapient Paradox' is the apparently unexplainable time delay of several ten thousand years following the arrival of *Homo sapiens* in Asia and Europe and before the introduction of impressive innovations with the agricultural revolution. Renfrew (2007) has suggested that the solution of the paradox has to do with changes in modes of thought that occurred with sedentism. According to Renfrew, this is a subject of study for cognitive archaeology where the final goal would be to understand the formation of the human mind. Several scholars, however, affirm that climatic change was crucial to such a revolution as it would have been very difficult to develop agriculture during the Palaeolithic. In other words, sedentism was not justified during the ice age, and that may be the solution to the paradox. It is widely accepted that climate variations were due to so-called orbital forcing, the slow periodic changes of orbital parameters of the Earth (known also as the Milankovitch theory). These and other astronomical effects on the climate are discussed along with the consequent impact on human evolution. The question then rises as to whether or not it is possible to adopt an 'astronomical' perspective instead of (or complementary to) the 'cognitive archaeological' one. Such would be possible by adopting a different point of reference (that is, from 'outside'), and a non-anthropocentric approach.

Keywords: cognitive archaeology, palaeoclimate, orbital forcing, solar activity

**Introduction**

The purpose of the present paper is to suggest the possibility of an astronomical perspective instead of (or complementary to) the cognitive archaeology perspective in the study of the human evolution that was proposed by Renfrew. He explained his approach clearly in papers and books (e.g. Renfrew 2007) and stressed the link between cognitive archaeology and cognitive science as an interdisciplinary science that includes neuroscience, psychology, and anthropology. His approach may be summarized as follows: 'the brain (mind) studies its own evolution'. On the 'Sapient Paradox', Renfrew (2007: 84) remarked:

> If the arrival [in Europe] of the new species, *Homo sapiens*, with its higher level of cognitive capacity […] was so significant, why did it take so long for the really impressive innovation seen in the accompanying agricultural revolution, to come about? What accounts for the huge gap from the first appearance of *Homo sapiens* in Europe 40,000 years ago (and earlier in western Asia) to the earliest agricultural revolution in western Asia and Europe of 10,000 years ago? This is a time lag of 30,000 years!

It is a central idea of Renfrew's work that the most decisive turn in prehistory came with an order-

---

of-magnitude increase in the varieties of engagement between humans and the material world, mediated by the use of symbols, which began with the development of sedentism at the beginning of Neolithic. While sedentism may have triggered the turn, what triggered the sedentism? According to Renfrew (2007: 147), 'most commentators, including Lewis Binford and Jacques Cauvin (e.g. Cauvin 2000), accept that climatic change was crucial (global warming, and the establishment of more stable conditions with fewer oscillations in temperature).' So, it might be concluded that it was climate change that triggered sedentism. However, what triggered climate change? Before trying to give an answer, it should be noted that apparently Renfrew (2007) didn't give much weight to the effects of climatic change.

**Orbital forcing**

*Effects on climate*

It is important to recall some of the widely accepted results of studies in paleoclimatology. The trend of the mean temperature difference estimated from deuterium in ice cores of Antarctica (e.g. Petit et al. 1999) suggests that climate of about 130 thousand and 240 thousand years ago was similar to the present (Figure 1). Going back further in time, other warm peaks with similar time scales can be seen, so the trend can be interpreted in terms of periodic phenomena (e.g. Kawamura et al. 2007). Analogous results have been obtained for all the various proxies that were analysed in the past forty years (e.g. from palynological sediments and oxygen isotopes in ocean sediments). According to paleoclimatologists these periodicities have an astronomical origin and are usually called orbital forcing (the Milankovitch theory). The slight changes in solar insolation due to slow changes in the eccentricity of the Earth's orbit, obliquity of the ecliptic, and precession of the Earth's orbit, gave rise to strong climate changes that can be deduced from the analysis of proxies. It is impressive that within paleoclimatology, there are no alternative hypotheses to the astronomical one, even if it does not exactly explain all data. Orbital forcing is at the basis of any interpretation of paleodata for the last millions years.

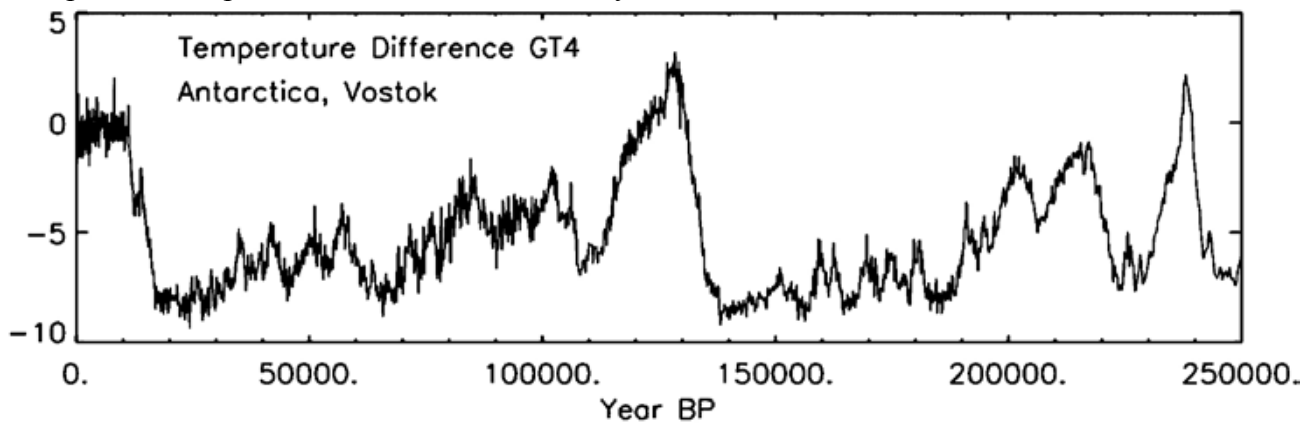

*Figure 1: Temperature difference GT4 in Antarctica according to Petit et al. (1999) against the years before present (BP). One should note the short duration of warm climate phases compared to the long duration of cold climate.*

From about ten thousand years ago to the present, the average temperature (not only in Antarctica) was rather stable, and the sea level significantly increased. Just few millennia before, the environment should have been completely different. The temperature was lower, and the sea level was very low, about 120 m below the present level (the late glacial maximum). Data on distribution of vegetation are extremely interesting (Figure 2). For example, on the mountains in southern Italy there were no deciduous woods as today, but there was a Siberian-like environment, a steppe (a lot of herbs), and the typical trees of cold climates. The very cold, dry, and unstable climate lasted at least fifty thousand years. Conversely, pollens suggest very stable and warm conditions during the last ten thousand years. This could be the explanation of the paradox. Indeed, it would have been rather difficult to imagine something like agriculture, at least at the latitudes above the tropic, during the long phase of cold climate. Actually, there was no reason for there to be a different economy from hunting and gathering during such a long period. Hence, there was no reason for sedentism, since it is generally supposed that the hunter-gatherer lifestyle was a nomadic one. Therefore one might conclude that it was orbital forcing that triggered the change in human evolution. As a logical consequence, it seems reasonable that in order to understand the human evolution, one has to know astronomy and its effects on climate.

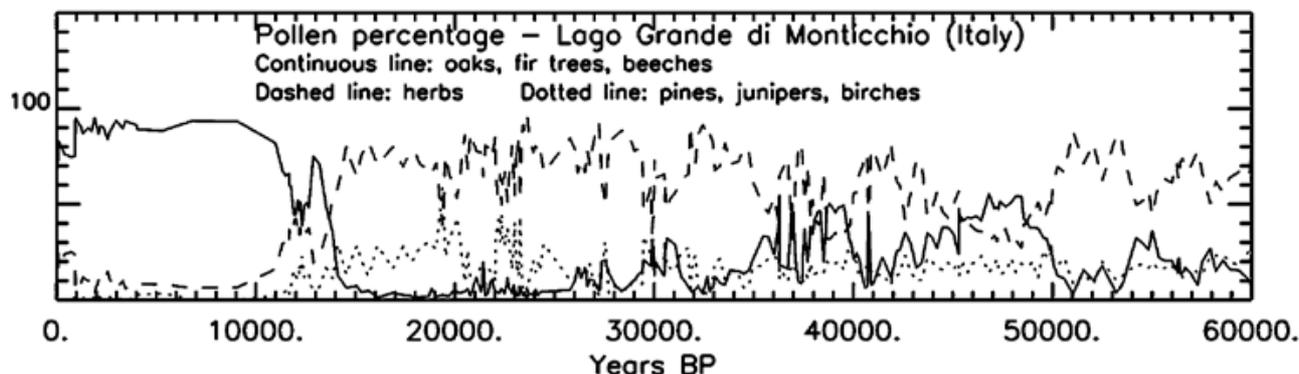

*Figure 2: Pollen percentage from the Lago Grande di Monticchio in southern Italy (Allen et al. 1999) plotted against the years before present (BP). Continuous line: deciduous trees of warm-temperate climate such as oaks and beeches, and fir trees; dashed line: herbs; dotted line: trees of cold climate such as pines, birches and junipers.*

*Caveat*

Of course the situation is not as clear-cut as depicted so briefly, and it is not possible to discuss it in detail here. Thus, some caveats are in order: *Homo sapiens*, our species, is supposed to have originated in Africa, and groups of *Homo sapiens* left from there about fifty thousand years ago (Mellars 2006). Some groups probably interbred with Neanderthals on their way to central Europe and with Denisovans in Asia (Gibbons 2011). The first farming economies appeared much later, about the ninth millennium BCE, and in the Near East (see e.g. Cauvin 2000). Recent data, however, from the New Guinea Highlands demonstrate exploitation of the endemic pandanus and yams in archaeological sites more than 40,000 years ago (Summerhayes et al. 2010). The sites contain stone tools thought to be used to remove trees, and this suggests that early inhabitants cleared forest patches to promote the growth of useful plants. In this case, however, *Homo sapiens* exploited wild (not domesticated) tubers and plants. This was not yet agriculture. In New Guinea, agriculture should have begun just 7000 thousand years ago with domesticated tubers, including taro, yams, sugarcane, and bananas. However, there are indications of forest clearance as early as 9000 years ago (Renfrew 2007: 61-62).

One could also wonder whether or not sedentism really began with climate change. Watkins (2010) pointed out that it probably began before, at least about 25,000 years ago (early Epipaleolithic) in the Near East (site of Ohalo II). Farming should have complemented rather than prompted the advent of permanent communities. There are also recent results regarding ancient pottery in China where vessels have been discovered dating back 20,000 years (Wu et al. 2012). There were probably hunters and gatherers living there in settlements. In any case, even taking into account these caveats, the basis of the evolution from Palaeolithic to Neolithic and the development of agriculture probably resulted from strong climate change due to orbital forcing.

**Other effects on climate**

Astronomical effects, however, also heavily affected the history of mankind at later times. For example, Wang et al. (2005) analysed sections of stalagmites of a cave in China. The general weakening of the Asian monsoon, i.e., progressively less rain during rainy seasons, was deduced by those authors for the last 9,000 years and corresponds with orbitally induced lowering of summer solar insolation (orbital forcing). Locally, during the Holocene Optimum Climate several thousand years ago, conditions were indeed wetter than present ones, for instance in the Sahara (Kuper and Kröpelin 2006). The general weakening trend is punctuated by eight further weak Asian Monsoon events, each lasting one to five centuries. Gupta et al. (2003) obtained analogous results from bio-sediments in the Arabian Sea. They correspond to so-called Bond events in the Northern emisphere detected in ice cores (Bond et al. 1997). The causes are not clear, but paleoclimatologists are thinking about astronomical effects such as changes in solar output or in the long period of lunar tides (1800 years) along with the relative atmospheric response of Earth. Some of these events produced strong droughts. Here, an event of about 4000 years ago should be noted. It has been detected in the sediments of the Gulf of Oman as a sudden increase of dust indicating strong aridification (de Menocal 2001; see also Brooks 2006). The consequences should have been dramatic for people whose lives depended on the East Asian, South-West Asian and West African monsoon regimes. Aridification was one of the most severe climatic events, and it very probably caused the collapse of the Old Kingdom in Egypt (presumably there were no more floods of the Nile), of the Akkadian Empire in Mesopotamia, and of the Neolithic civilization in China. These are just some examples of the terrible effects that astronomical parameters seem to have had on climate and on timescales of centuries and millennia. The last Bond event corresponds to the so-called little ice age of three centuries ago which is usually ascribed to lower solar activity as indicated by the Maunder minimum, a lack of solar spots. The physical mechanism that has been proposed is related to the weak solar magnetic field during a lower activity phase; the field would be no more able to effectively shield the Solar System from the galactic cosmic rays. Their effect would be to change the cloud covering conditions in the atmosphere (for a critical review about the astronomical impacts on climate, see Bailer-Jones 2009).

**An astronomical perspective**

Astronomy therefore must be taken into account in order to understand the evolution and the history of mankind, and this suggests the possibility of adopting an 'astronomical perspective'. It would consist of the study of evolution of mankind (and also of the history of Earth and of the evolution of life on Earth) as seen from 'outside', not from 'inside'. Is it possible to adopt a different point of reference from that of cognitive archaeology and to study human evolution with a 'non-anthropocentric' approach? This may be more understandable by scholars that have an education in astronomy and/or astrophysics. In astronomy, the Earth is studied as a planet in the same way that other Solar System objects are studied, and it is not a special place apart from the fact that it is the only place where the peculiar phenomenon of life has been observed thus far. On the other hand, the approach of cognitive archaeology is the study of mankind's evolution as seen so to speak from 'inside' the mind itself. This seems an essentially anthropocentric approach.

    For scholars that received a humanistic education, the astronomical perspective may not be as evident as for astronomers. Therefore, in order to help in understanding, an analogy from science fiction may be used. Imagine a large spaceship, even larger than Enterprise of Star Trek. It has a very long-term mission that will last about twenty years, and it consists in the exploration of some regions of the Galaxy. There are several hundred people on board. The spaceship is thus a small travelling town. During the long mission, children are born and are educated on board. One of them is a young scholar interested in the study of possible (if any) civilizations in planetary systems of

the Galaxy. To him, the Sun is not at all a special star, and the Solar System is just one of the systems with a planet harbouring life. When at the end of the mission, he has the opportunity to land on the planet Earth and to visit it, he does not feel himself an earthling. He comes from space and feels himself to be an alien. His approach will be the study of the human population on Earth as seen from 'outside'. In a certain sense, this is the approach of a present day astronomer or astrophysicist.

Generally, an 'astronomical' perspective should allow a better global vision of the Earth's history. A better synthesis would be possible of what happened on this planet if an astronomical framework were used. Of course, this could be just a professional bias. Therefore, comments on this idea are welcome.